\documentclass{article}

\usepackage{amssymb}
\usepackage{graphicx}
\begin{document}

\begin{center}

{\Large\bf CONSTRAINING A DOUBLE COMPONENT\\[5PT]
DARK ENERGY MODEL USING\\[5PT]
SUPERNOVA TYPE IA DATA\\[5PT]}
\medskip

J.C. Fabris\footnote{e-mail: fabris@cce.ufes.br}, S.V.B.
Gon\c{c}alves\footnote{e-mail: sergio@cce.ufes.br}, \medskip

Departamento de F\'{\i}sica, Universidade Federal do Esp\'{\i}rito Santo,
29060-900, Vit\'oria, Esp\'{\i}rito Santo, Brazil
\medskip

Fabr\'{\i}cio Casarejos \footnote{e-mail: fabll@dft.if.uerj.br} and
Jaime F. Villas da Rocha\footnote{e-mail: roch@dft.if.uerj.br}

\medskip

Instituto de F\'{\i}sica, Universidade Estadual do Rio de Janeiro,
20550-900, Rio de Janeiro, Brazil

\end{center}

\begin{abstract}

A two-component fluid representing dark energy is studied. One of
the components has a polytropic form, while the other has a
barotropic form. Exact solutions are obtained and the cosmological
parameters are constrained using supernova type Ia data. In general,
an open universe is predicted. A big rip scenario is largely preferred, but the dispersion in the
parameter space is very high. Hence, even if scenarios without future
singularities can not be excluded with the allowed range of
parameters, a phantom cosmology, with an open spatial section, is a general prediction of the model.
For a wide range of the equation of state parameters there is an asymptotic de Sitter phase.

\end{abstract}

Pacs numbers: 98.80.-k, 98.80.Es

\section{Introduction}

Several cosmological observables indicate that the present
Universe is in a state of accelerated expansion. The first
evidence in this sense came in the end of the last decade, when
two independent observational projects \cite{SN}, using type
Ia supernovae luminosity distance-redshift relation, provided an
estimation of the deceleration parameter $q = - a\ddot a/\dot
a^2$. With a catalogue of about $50$ type Ia supernova with low
and high redshifts, the analysis revealed a negative $q$,
indicating an accelerated Universe. Today, more than $300$ type Ia
supernovae have been identified, with high redshift, and
the conclusion that $q$ is negative remains \cite{tonry}. Since
then, there has been an extensive discussion on the quality of the
data. This led to a restricted sample of $157$ supernova, called
the "gold sample" \cite{riess}. A more recent survey led to the
so-called "legacy sample", of about $100$ supernova, with high
quality data \cite{astier}. Even if the precise estimation of the
cosmological parameters, like the matter density, Hubble
parameter, etc, depends quite strongly on the choice of the sample,
the conclusion that the Universe is accelerating has remained.
Hence, a large part of the community of cosmologist accepts the
present acceleration of the Universe as a fact.
\par
A combination data from CMB anisotropies of the cosmic microwave
background radiation \cite{verde}, large scale structure
\cite{tegmark1} and type Ia supernovae data \cite{colistete},
indicates an almost flat Universe, $\Omega_T \sim 1$ and a matter
(zero effective pressure) density parameter of order $\Omega_m
\sim 0.3$ \cite{tegmark2}. Since an accelerated expansion can be
driven by a repulsive effect, which can be provided by an exotic
fluid with negative pressure, it has been concluded that the
Universe is also filled by an exotic component, called dark
energy, with density parameter $\Omega_c \sim 0.7$. This exotic
fluid leads to an accelerated expansion, remaining at same time
smoothly distributed, not appearing in the local matter
clustering.
\par
The first natural candidate to represent dark energy is a cosmological
constant, which faces, however, many well-known problems. More recently,
other candidates have been studied in the literature:
quintessence, k-essence, Chaplygin gas, among many others. For a
review of these proposals, see reference \cite{sahni}. There are
also claims that a phantom field (fields with a large negative
pressure such that all energy conditions are violated) leads to
the best fit of the observational data \cite{fantasma}. A phantom
field implies a singularity in a finite future proper time, which
has been named big rip, where density and curvature diverge.
This is of course an undesirable feature, but more detailed
theoretical and observational analyses must be made in order to
verify this scenario. Some authors state, based on considerations
about the evaluation of the cosmological parameters, that there is
no such {\it phantom menace} \cite{pad}. This is still an
object of debate.
\par
Most of the studies made until now lay on the assumption of a
simple relation between pressure and density expressed generically,
in a hydrodynamical representation,
by $p = w\rho^\alpha$. Quintessence, like others dark energy candidates, imply that $w$ varies with the redshift,
not being a constant. Chaplygin gas models (generalized or not) \cite{chaplygin} imply a general value for $\alpha$, but typically negative, and $w < 0$.
Phantom fields could be
represented by $\alpha = 1$, $w < - 1$. Due to the high speculative nature of the dark energy component, many possibilities
have been considered in the literature, both from fundamental or phenomenological point of views.
\par
In the present work we intend to exploit a more generic relation
between pressure and density with respect to those cases normally
considered. The main idea is to use a double component equation of
state. The relation between pressure and density may be written as
\begin{equation}
\label{eos}
p_e = - k_1\rho_c^\alpha - k_2\rho_c \quad .
\end{equation}
where $\alpha$, $k_1$ and $k_2$ are constants, and the subscript
$c$ indicates that such relation concerns the dark energy
component of the matter content of the Universe. Let us call the
component labelled by $k_1$ as the {\it polytropic component}, and
that one labelled by $k_2$ as the {\it barotropic component}. This
kind of equation of state has been, for example, studied in a
theoretical sense in reference \cite{indianos}.
\par
The equation of state (\ref{eos}) may be also seen as a
realisation of the so-called {\it modified Chaplygin gas}
\cite{benaoum,chakra1,chakra2}. The usual Chaplygin gas model,
generalised or not, has been introduced in order to obtain an
interpolation between a matter dominated era and a de Sitter phase
\cite{chaplygin}. The modified Chaplygin gas model allows to
obtain an interpolation between, for example, a radiative era and
a $\Lambda CDM$ era. In general, in this case, a negative value
for $\alpha$ is considered. But, as it will be seen later, even
for a positive value of $\alpha$, such an interpolation is
possible. From the fundamental point of view, the equation of
state (\ref{eos}) can be obtained in terms of self-interacting
scalar field. In references \cite{benaoum,chakra1,chakra2} a
connection with the rolling tachyon model has been established.
This allows to consider the equation of state (\ref{eos}) as a
phenomenological realisation of a string specific configuration.
\par
In reference \cite{alcaniz1}, a structure similar to ({\ref{eos})
has been analysed, using observational data, but fixing $k_2 = 1$,
with the conclusion that the fitting of the supernova data are
quite insensitive to the parameter $\alpha$. Here, we follow
another approach: we will fix $\alpha = 1/2$, leaving $k_2$ free.
This has the advantage of leading to explicit analytical
expressions for the evolution of the Universe. Moreover, and
perhaps more important, this may lead to interesting scenarios
where, for example, the Universe evolves asymptotically as in a de
Sitter phase, even if the equation of state is not characteristic
of the vacuum state, $p = - \rho$.
\par
We will test the equation of state (\ref{eos}) against type Ia
supernovae data. We will span a four dimensional phase
space, using as free parameters the dark matter density parameter
$\Omega_{m0}$, the exotic fluid density parameter $\Omega_{c0}$ (or alternatively, the
curvature parameter $\Omega_{k0}$), the Hubble parameter $H_0$, and the equation of
state parameter $k_1$ or $k_2$. The subscript $0$ indicates that all these quantities are evaluated today.
Using the gold sample, we will
show that the preferred values indicate $k_2 \sim 6$, $\Omega_{m0}$ and $\Omega_{c0}$
$\sim 0.3$, and $H_0 \sim 67$. An open universe is a general prediction for this model. A phantom behaviour is
largelly
favoured. However, the dispersion is very high, and an asymptotic
cosmological constant phase can not be discarded.
\par
The use of other observables, like
the spectrum of the anisotropy of the cosmic microwave background radiation (CMB) and
the matter power spectrum, can in principle restrict more severely the parameter space.
However, we postpone this evaluation to a future study because, in both cases, a perturbative analysis of the
model is necessary. In this case we must replace the hydrodynamical representation
presented above by a fundamental description of the fluid, for example, in terms of self-interacting scalar
fields. We note {\it en passant} that the hydrodynamical representation employed here may lead, at perturbative level, to instabilities
at small scales due to a imaginary effective sound velocity, instabilities that can be avoided with
a fundamental representation \cite{jerome}. There are many different ways to implement this more fundamental description, which
can lead to different results. The supernova data, on the other hand, test
essentially the background, which is somehow independent of the description of the fluid.
\par
The paper is organised as follows. In the next section, we obtain
some analytical expressions for the evolution of the Universe, and
derive the luminosity distance relation for the model. In section
$3$, we make the comparison between the theoretical model and the
observational data. In section $4$, we present our conclusions.

\section{The evolution of the Universe}

Let us consider the equations of motion when the exotic fluid
given by the equation of state (\ref{eos}) dominates the matter
content of the Universe. The Friedmann's equation and the conservation of the
energy-momentum tensor read,
\begin{eqnarray}
\label{em1a}
\biggr(\frac{\dot a}{a}\biggl)^2 + \frac{k}{a^2} &=& \frac{8\pi G}{3}\rho_c \quad , \\
\label{em2a}
\dot\rho_c + 3\frac{\dot a}{a}(\rho_c + p_c) &=& 0 \quad ,
\end{eqnarray}
where $k$ is the curvature of the spatial section.
Inserting equation (\ref{eos}), with $\alpha = 1/2$, in equation (\ref{em2a}), it comes
out that the exotic fluid density depends on the scale factor as
\begin{equation}
\label{ef}
\rho_c = \frac{1}{\beta^2}\biggr[k_1 + v_0a^{-\frac{3}{2}\beta}\biggl]^2 \quad ,
\end{equation}
where $\beta = 1 - k_2$. Introducing this result in equation
(\ref{em1a}), it is possible to obtain an explicit solution for
the scale factor when $k = 0$:
\begin{equation}
a = a_0\biggr\{\exp\biggr[k_1 M\,t\biggl] -
c_0\biggl\}^{\frac{2}{3\beta}} \quad ,
\end{equation}
where $a_0$ and $c_0$ are integration constants. The constants obey the relations
\begin{equation}
M = \sqrt{6\pi G} \quad , \quad c_0 =
\frac{v_0}{k_1}a_0^{-\frac{3}{2}\beta} \quad .
\end{equation}
This solution can always represent an expanding Universe, with an
initial singularity. Moreover, when $k_2 > 1$, the density goes to
infinity as the scale factor goes to infinity, in a finite proper
time, characterising a big rip. However, if $k_2 < 1$, the
expansion lasts forever, and becomes asymptotically de Sitter even
if $k_2 \neq 1$ (the strict cosmological constant case). Initially, the scale
factor behaves as in the pure barotropic case with $p = - k_2\rho$.
\par
The particular case where $k_2 = 1$, but with free $\alpha$, has
been analysed in reference \cite{alcaniz1}. For our case, fixing
$\alpha = 1/2$ and $k_2 = 1$, the relation between density and
scale factor becomes,
\begin{equation}
\rho_c = \biggr(\frac{3}{2}k_1 - 1\biggl)\ln a \quad .
\end{equation}
\par
If we consider the dynamics of a universe driven by the exotic
fluid defined by equation (\ref{eos}) and pressureless matter, the
equations of motion are given by,
\begin{eqnarray}
\label{em1}
H^2 &=& \biggr(\frac{\dot a}{a}\biggl)^2 = \frac{8\pi G}{3}(\rho_m + \rho_c) - \frac{k}{a^2} \quad ,\\
\label{em2}
\dot\rho_m + 3\frac{\dot a}{a}\rho_m &=& 0 \quad , \\
\label{em3}
\dot\rho_c + 3\frac{\dot a}{a}(\rho_c + p_c) &=& 0 \quad ,
\end{eqnarray}
where $p_c$ is given by equation(\ref{eos}). The conservation
equations (\ref{em2},\ref{em3}) can be integrated, again for $\alpha = 1/2$, leading to the
relation (\ref{ef}) and $\rho_m = \rho_{m0}/a^3$.
In this case, it does not seem possible to obtain a closed expression for the scale factor
in terms of the cosmic time $t$ as before. However, the inclusion of the pressureless component
is essential in order to take into account the effects of the baryons in the determination of the
allowed range for the parameters of the model using the supernova data, as it will be done in
the next section.

\section{Fitting type Ia supernovae data}

As time goes on, more and more high redshift type Ia supernovae
are detected. Today, about 300 high $z$ SN Ia have been reported.
However, there are many discussions on the quality of these data.
A "gold sample", with the better SN Ia data, with a number of
$157$ SN, has been proposed \cite{riess}. More recently, the {\it Supernova
Legacy Survey} (SNLS) was made public, containing around 100 SNIa
\cite{astier}. In this work, we will use the gold sample. This
will allows us to compare our results with previous ones using a
similar method, but with different models \cite{colistete}.
\par
From now on, we will normalise the scale factor, making it equal
to one today: $a_0 = 1$. Hence, the relation between the scale
factor $a$ and the redshift $z$ becomes $1 + z = 1/a$. In order to
compare the observational data with the theoretical values, the
fundamental quantity is the luminosity distance
\cite{weinberg,coles}, given by
\begin{equation}
D_L = (1 + z)r \quad ,
\end{equation}
where $r$ is the comoving radial position of the supernova. For a flat Universe, the comoving radial coordinate is given by
\begin{equation}
r = \int_0^z\frac{dz'}{H(z')} \quad .
\end{equation}
Using equation (\ref{em1}), with the expressions for the exotic
and pressureless fluid in terms of $a$, converted to relations for
those components in terms of the redshift $z$, we obtain the
dependence of the Hubble parameter in terms of $z$. Hence, the
final expression for the luminosity distance is, for our model
with $k_2 \neq 1$,
\begin{eqnarray}
& &D_L = \frac{c(1 + z)}{H_0}\int_0^z\biggr\{\Omega_{m0}(1+z')^3 + \Omega_{k0}(1 + z')^2
\nonumber\\
&+&
\Omega_{c0}\biggr[\biggr(1-\frac{k_1}{1-k_2}\biggl)(1+z')^{3\frac{(1-k_2)}{2}}+\frac{k_1}{1-k_2}\biggl]^2\biggl\}^{-1/2}dz'
\quad .
\end{eqnarray}
For the case $k_2 = 1$, the luminosity distance is given by
\begin{eqnarray}
D_L = \frac{c(1 + z)}{H_0}\int_0^z\frac{dz'}{\sqrt{\Omega_{m0}(1 +
z')^3 + \Omega_{k0}(1 + z')^2 + \Omega_{c0}\biggr\{1 - \frac{3}{2}k_1\ln(1 +
z')\biggl\}^2}} \quad .
\end{eqnarray}
In the expressions above, $H_0$ is the Hubble parameter today,
which can be parametrized by $h$, such that $H_0 =
100\,h\,km/s\cdot Mpc$. The parameter $k_1$ has been redefined as
$k_1/\sqrt{\rho_{c0}} \rightarrow k_1$, $\rho_{c0}$ being the
exotic component density today. This redefinition is made in order
to obtain a dimensionless parameter $k_1$. Moreover, $\Omega_{m0}
= 8\pi G\rho_{m0}/(3H_0^2)$, $\Omega_{c0} = 8\pi
G\rho_{c0}/(3H_0^2)$ and $\Omega_{k0} = - k/H_0^2$.
\par
The comparison with the observational data is made by computing
the distance modulus, defined as
\begin{equation}
\mu_0 = 5\log\biggr(\frac{D_L}{Mpc}\biggl) +\,25 \quad ,
\end{equation}
which is directly connected with the difference between the
apparent and absolute magnitudes of the supernovae. The quality of
the fitting is given by:
\begin{equation}
\chi^2 = \sum_i\frac{(\mu_{0i}^o - \mu_{0i}^t)^2}{\sigma^2_{0i}}
\quad ,
\end{equation}
where $\mu_{0i}^o$ and $\mu_{0i}^t$ are the observed and calculated distance moduli for the
$ith$ supernova, respectively, while $\sigma^2_{0i}$ is the
error in the observational data, taking already into account the effect of the dispersion due to the peculiar velocity.
\par
In principle, the model contains five free parameters: $k_1$,
$k_2$, $H_0$, $\Omega_{c0}$ and $\Omega_{m0}$. We will work in a four dimensional
phase space: for each value of $k_1$, we will vary the other four
parameters. Thus, the $\chi^2$ function will depend on four
parameters, $\Omega_{c0}$, $\Omega_{m0}$, $H_0$ and $k_2$. The probability
distribution is then given by
\begin{equation}
\label{prob} P(\Omega_{c0},\Omega_{m0},H_0,k_2) = A\,e^{-
\frac{\chi^2(\Omega_{c0},\Omega_{m0},H_0,k_2)}{2\sigma}} \quad ,
\end{equation}
where $A$ is a normalisation constant and $\sigma$ is directly
related to the confidence region. The graphics and the parameter estimations
were made using the software {\bf BETOCS} \cite{betocs}.

\begin{figure}[!t]
\begin{minipage}[t]{0.5\linewidth}
\includegraphics[width=\linewidth]{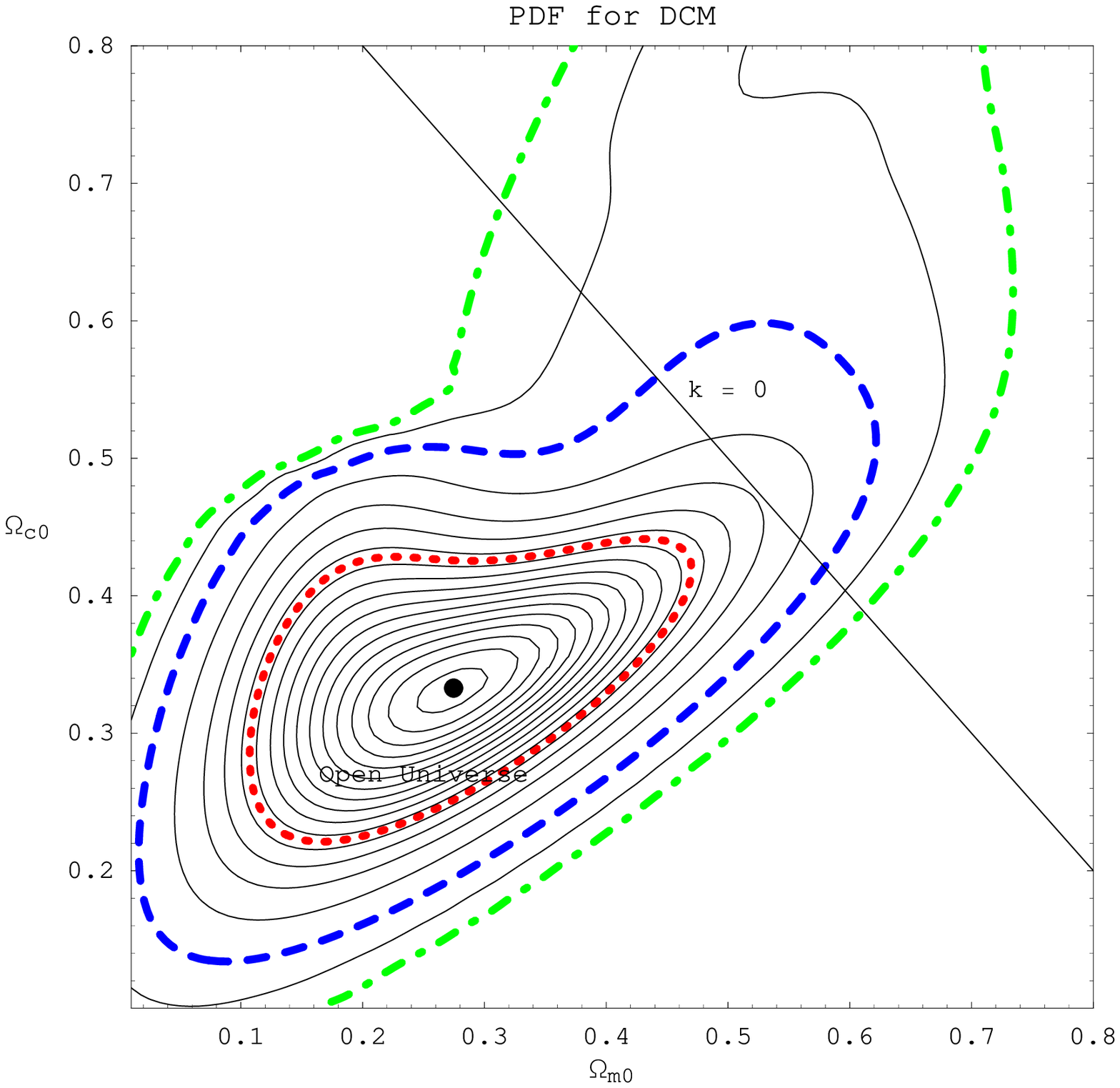}
\end{minipage} \hfill
\begin{minipage}[t]{0.5\linewidth}
\includegraphics[width=\linewidth]{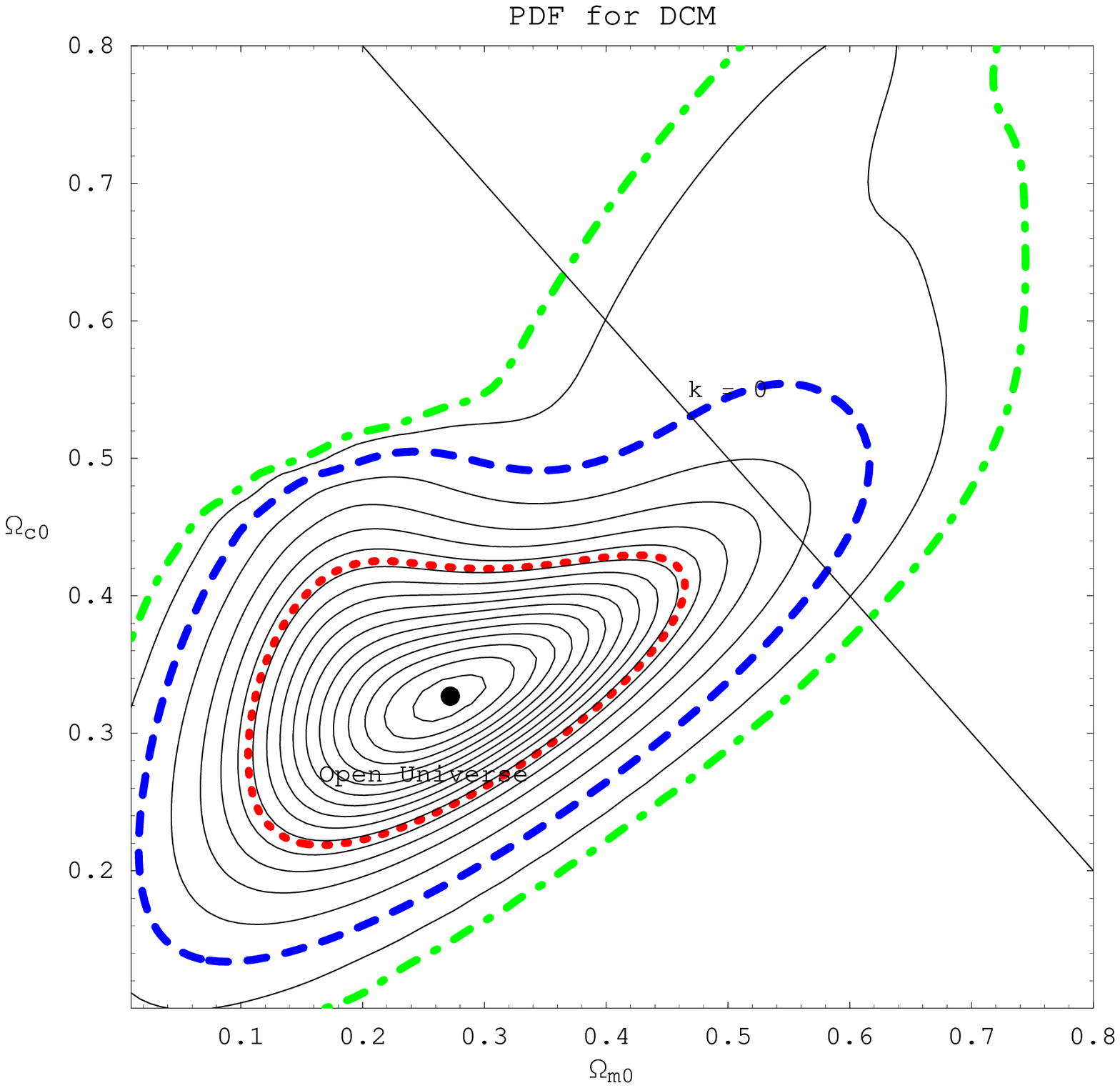}
\end{minipage} \hfill
\caption{{\protect\footnotesize The plots of the joint PDF as
function of $(\protect \Omega_{c0},\Omega_{m0})$ for the two component
model, for $k_1 = 0$ and $1.0$. The
joint PDF peak is shown by the large dot, the confidence regions of $1\,\protect\sigma $ ($68,27\%$) by the dotted line, the
$2\,\protect\sigma $ ($95,45\%$) in dashed line and the
$3\,\protect\sigma $ ($99,73\%$) in dashed-dotted line.}}
\end{figure}

\begin{figure}[!t]
\begin{minipage}[t]{0.5\linewidth}
\includegraphics[width=\linewidth]{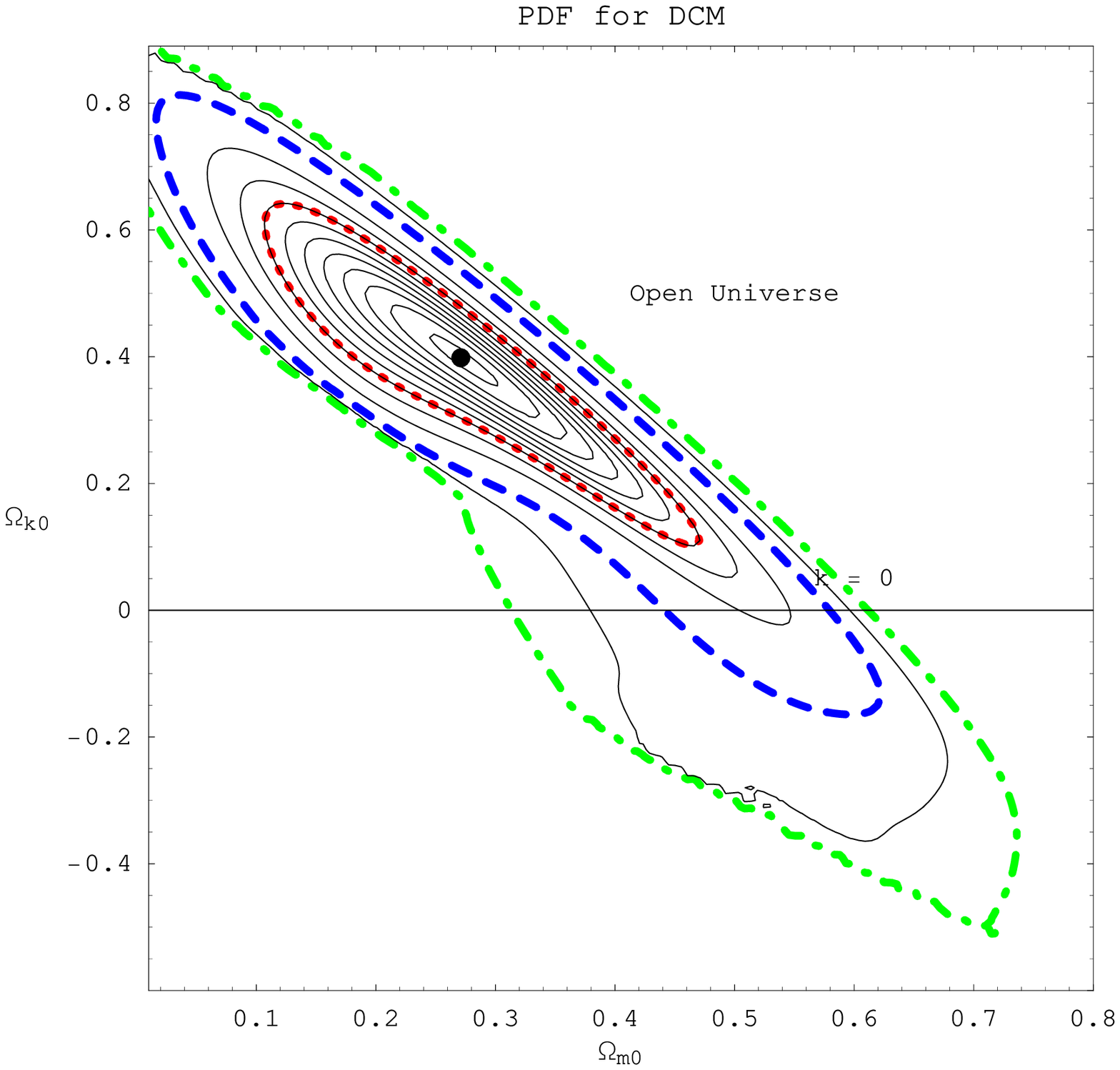}
\end{minipage} \hfill
\begin{minipage}[t]{0.5\linewidth}
\includegraphics[width=\linewidth]{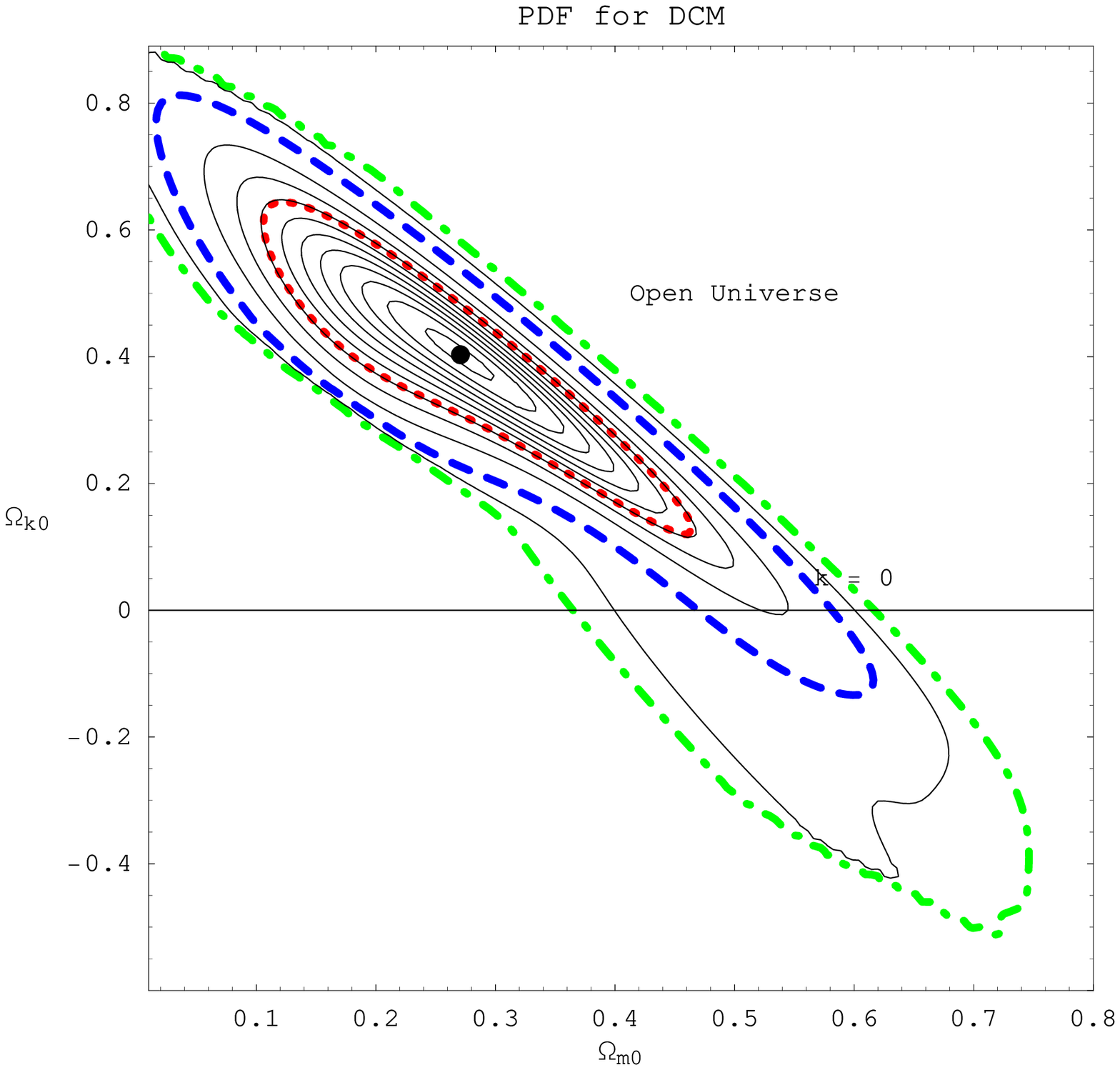}
\end{minipage} \hfill
\caption{{\protect\footnotesize The plots of the joint PDF as
function of $(\protect \Omega_{k0},\Omega_{m0})$ for the two component
model, for $k_1 = 0$ and $1.0$. The
joint PDF peak is shown by the large dot, the confidence regions of $1\,\protect\sigma $ ($68,27\%$) by the dotted line, the
$2\,\protect\sigma $ ($95,45\%$) in dashed line and the
$3\,\protect\sigma $ ($99,73\%$) in dashed-dotted line.}}
\end{figure}

\begin{figure}[!t]
\begin{minipage}[t]{0.5\linewidth}
\includegraphics[width=\linewidth]{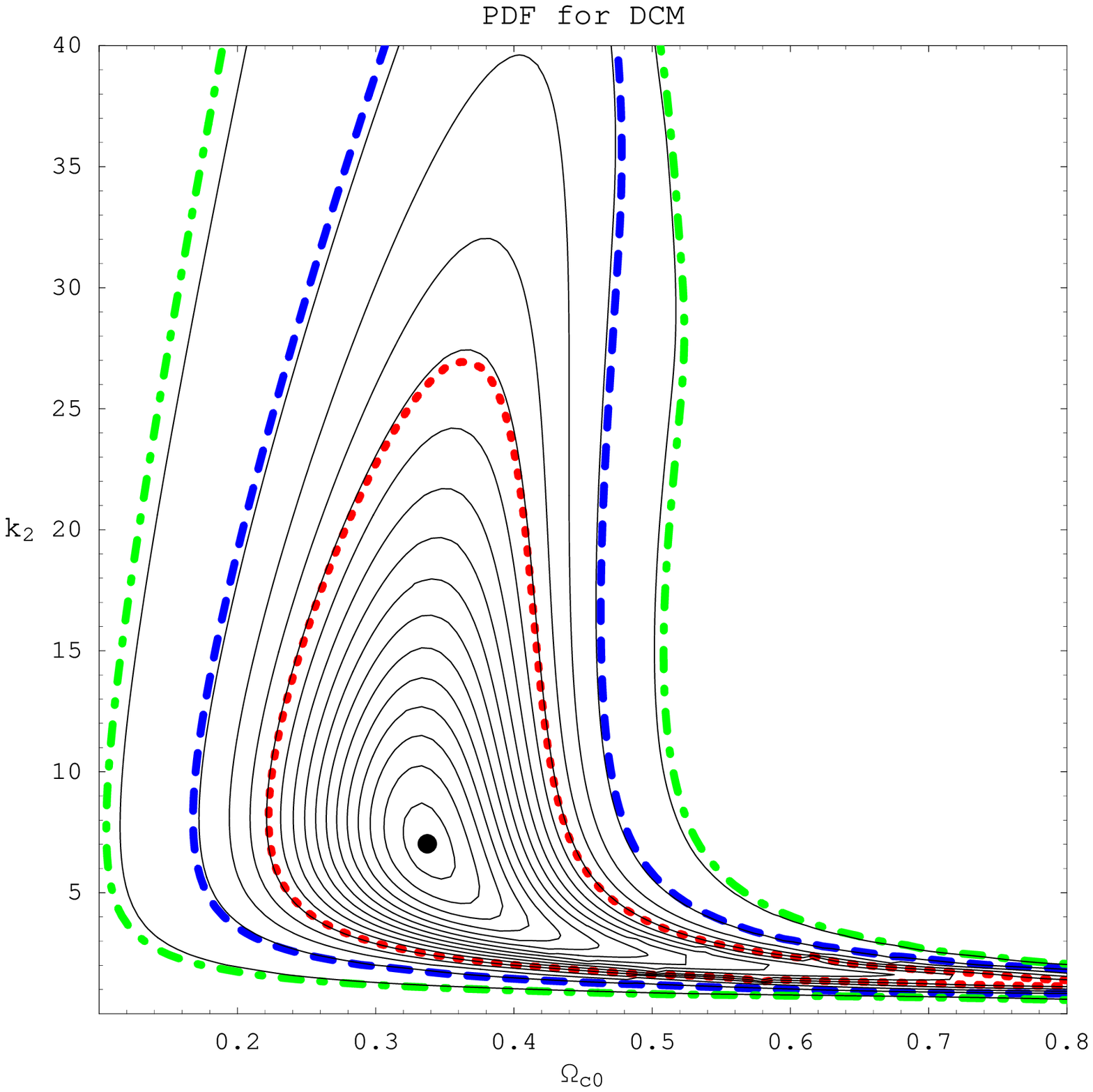}
\end{minipage} \hfill
\begin{minipage}[t]{0.5\linewidth}
\includegraphics[width=\linewidth]{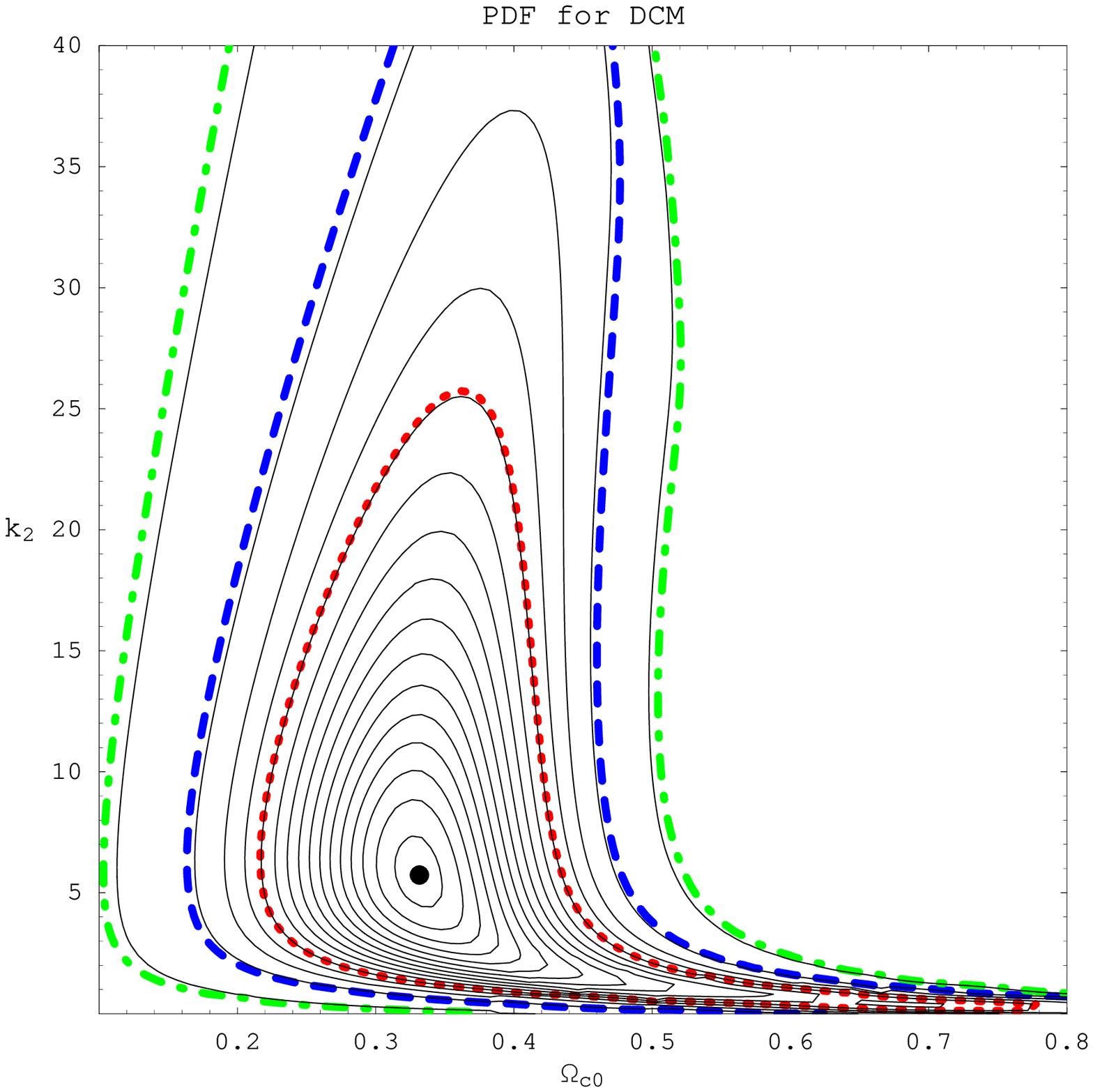}
\end{minipage} \hfill
\caption{{\protect\footnotesize The plots of the joint PDF as
function of $(\protect k_2, \Omega{c0})$ for the two component model, for
$k_1 = 0$ and $1.0$. The
joint PDF peak is shown by the large dot, the confidence regions of $1\,%
\protect\sigma $ ($68,27\%$) by the dotted line, the
$2\,\protect\sigma $ ($95,45\%$) in dashed line and the
$3\,\protect\sigma $ ($99,73\%$) in dashed-dotted line.}}
\end{figure}

\begin{figure}[!t]
\begin{minipage}[t]{0.5\linewidth}
\includegraphics[width=\linewidth]{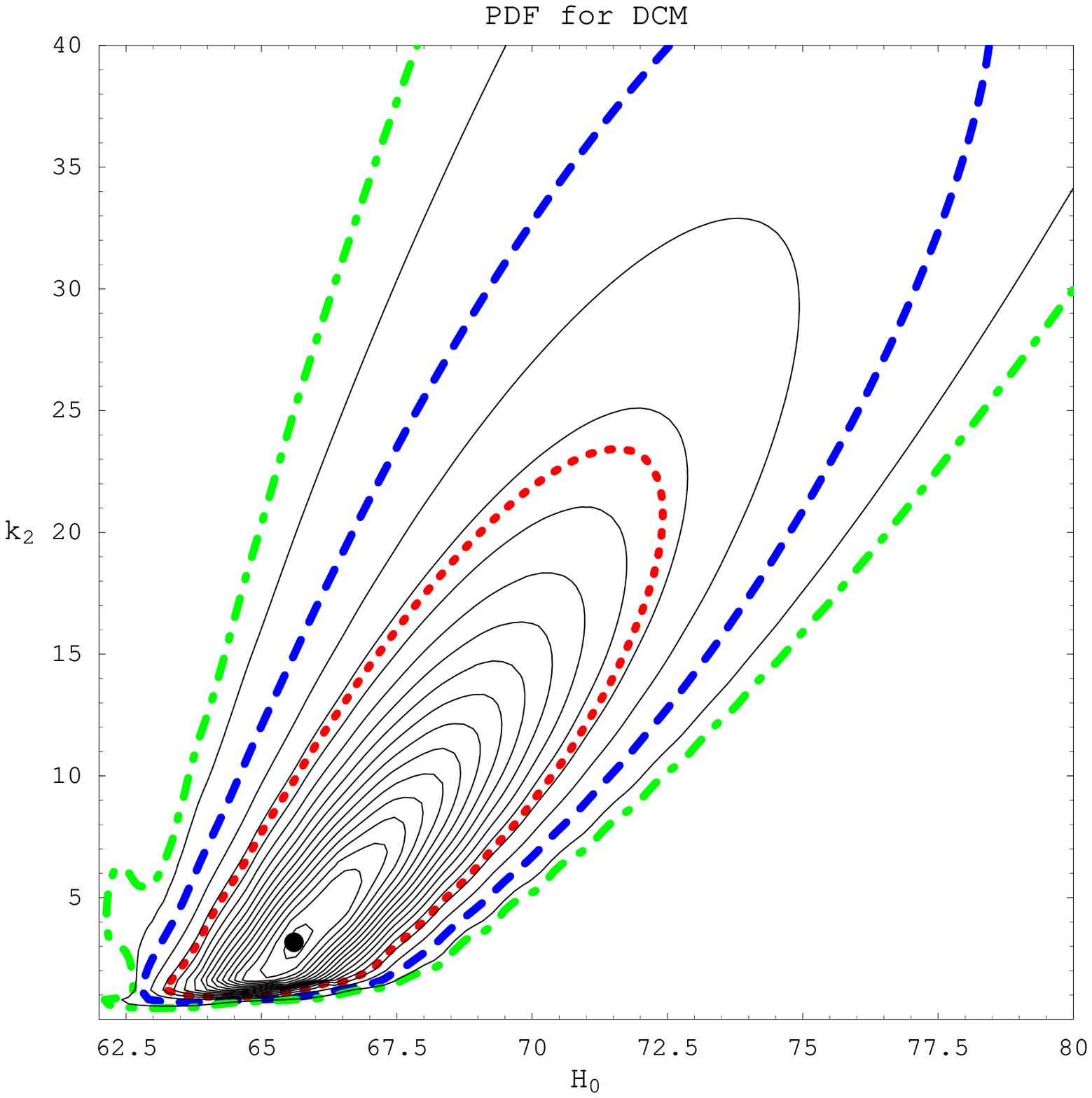}
\end{minipage} \hfill
\begin{minipage}[t]{0.5\linewidth}
\includegraphics[width=\linewidth]{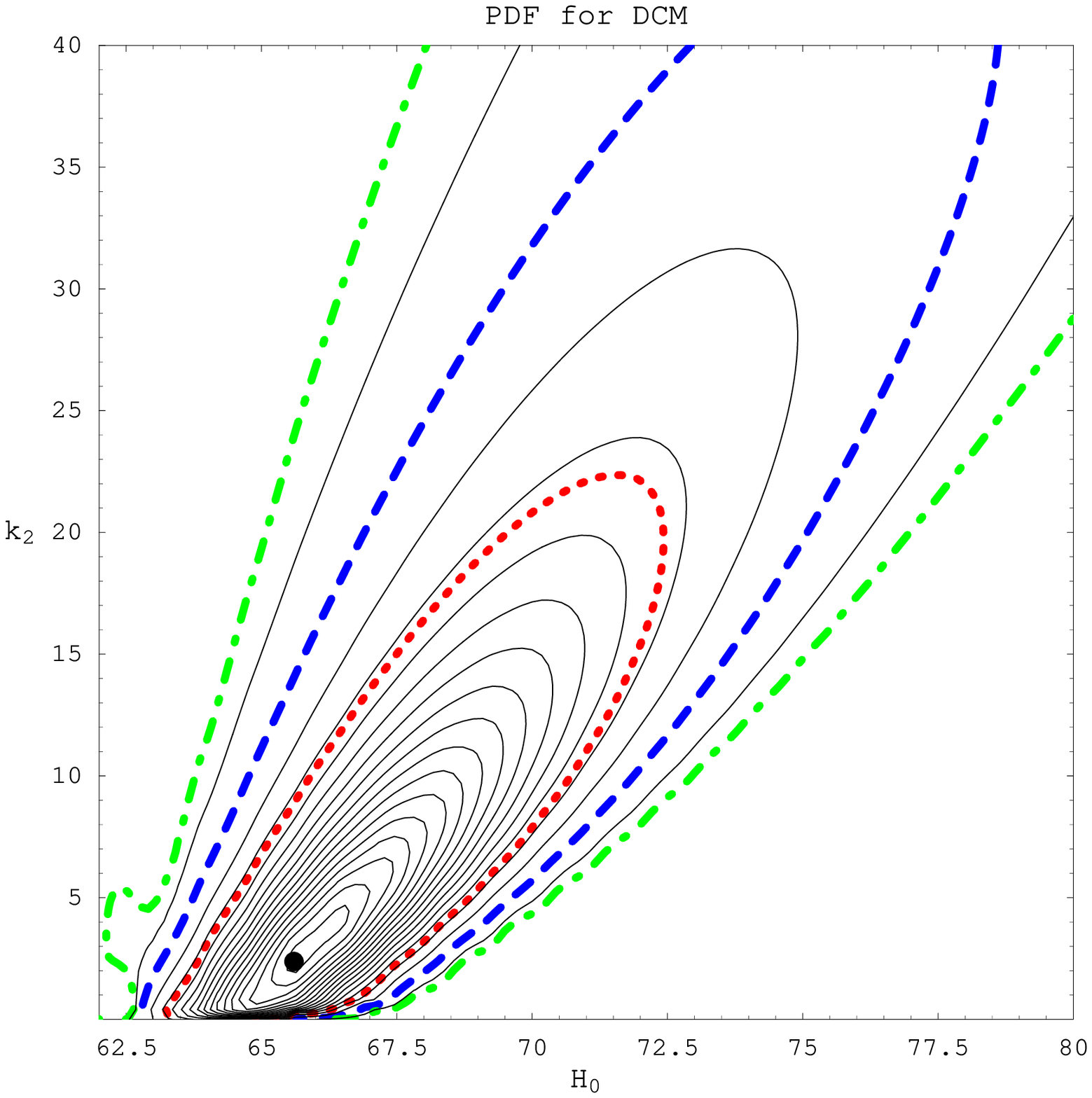}
\end{minipage} \hfill
\caption{{\protect\footnotesize The plots of the joint PDF as
function of $(\protect k_2,H_0)$ for the two component
model, for $k_1 = 0$ and $1.0$. The
joint PDF peak is shown by the large dot, the confidence regions of $1\,%
\protect\sigma $ ($68,27\%$) by the dotted line, the
$2\,\protect\sigma $ ($95,45\%$) in dashed line and the
$3\,\protect\sigma $ ($99,73\%$) in dashed-dotted line.}}
\end{figure}
\par
The multidimensional plot of the probability distribution is not,
in general, the best way to have an overview of the results.
However, we can construct a two dimensional probability
distribution, integrating in two of the parameters. In figure $1$,
we display the Probability Density Function, PDF, after
integrating the distribution (\ref{prob}) in the variables $H_0$
and $k_2$: a two-dimensional probability distribution is obtained
for the variables $\Omega_{m0}$ and $\Omega_{c0}$, and with $k_1 =
0$ and $1.0$. The plots show the confidence regions at $1\sigma$
($68\%$), $2\sigma$ ($95\%$) and $3\sigma$ ($99\%$) levels. This
two-dimensional probability distribution reveals that the matter
density parameters have their higher probability around
$\Omega_{m0}, \Omega_{c0} \sim 0.3$. An open model is clearly
preferred. This is also evident in figure $2$, where the
two-dimensional probability distribution for $\Omega_{k0}$ and
$\Omega_{m0}$ is shown. Such preference for an open model contrast
strongly with a similar analysis for the $\Lambda CDM$ and
Chaplygin gas models, for which a closed universe is clearly
preferred \cite{colistete}. As the value of $k_1$ grows, a low
density universe becomes more favoured, as it can be seen in table
$1$. In this table, we include also the case $k_1 = 0.5$ to show more
explicitely that the parameter estimations change very slightly with $k_1$.
\par
In figure $3$, the two-dimensional PDF is displayed when we
integrate on $\Omega_{m0}$ and $H_0$, remaining with $k_2$ and $\Omega_{c0}$. As in the preceding case, a low density parameter for the dark energy component is preferred.
What is specially interesting is that the values of $k_2$ larger than $1$ and until around $40$ have higher probabilities.
This means that that a phantom scenario is clearly favoured. The peak of probability for $k_2$ occurs near $5-6$. As the value of $k_1$ increases,
the peak probability for $k_2$ occurs at a lower value.
\par
In figure $4$, the two-dimensional PDF is displayed when we
integrate on $\Omega_{m0}$ and $\Omega_{c0}$, obtaining a two-dimensional graphic for
$k_2$ and $H_0$. It is interesting to note, now, that regions around $H_0 \sim 70 km/Mpc.s$ are preferred.
This may reconcile the estimations
obtained using supernova with those obtained using CMB and matter clustering, which indicates $H_0$ around $72 km/Mpc.s$ \cite{tegmark1, wmap}.
The prefered value for $H_0$ depends very little on $k_1$.

\begin{figure}[!t]
\begin{minipage}[t]{0.5\linewidth}
\includegraphics[width=\linewidth]{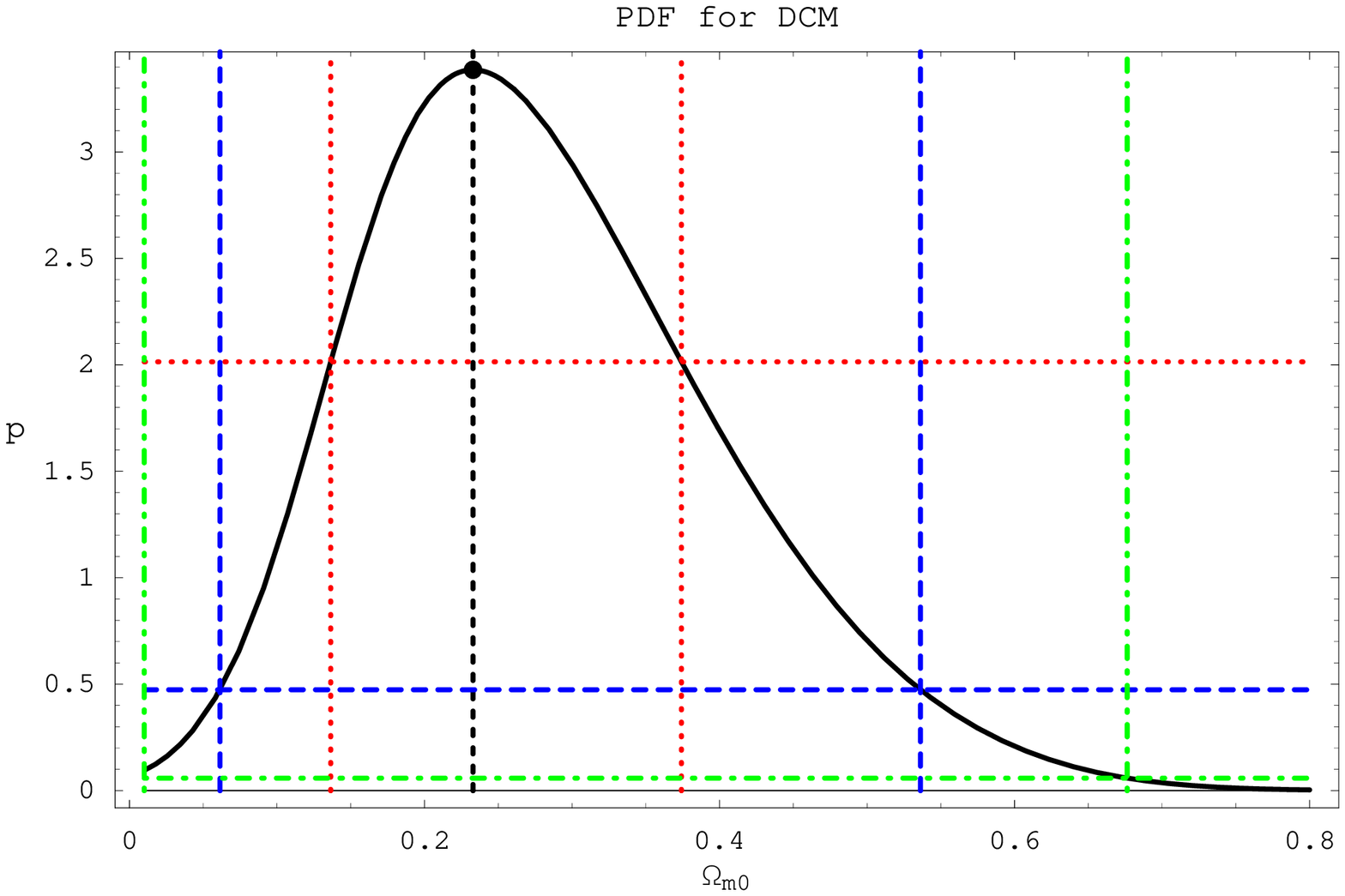}
\end{minipage} \hfill
\begin{minipage}[t]{0.5\linewidth}
\includegraphics[width=\linewidth]{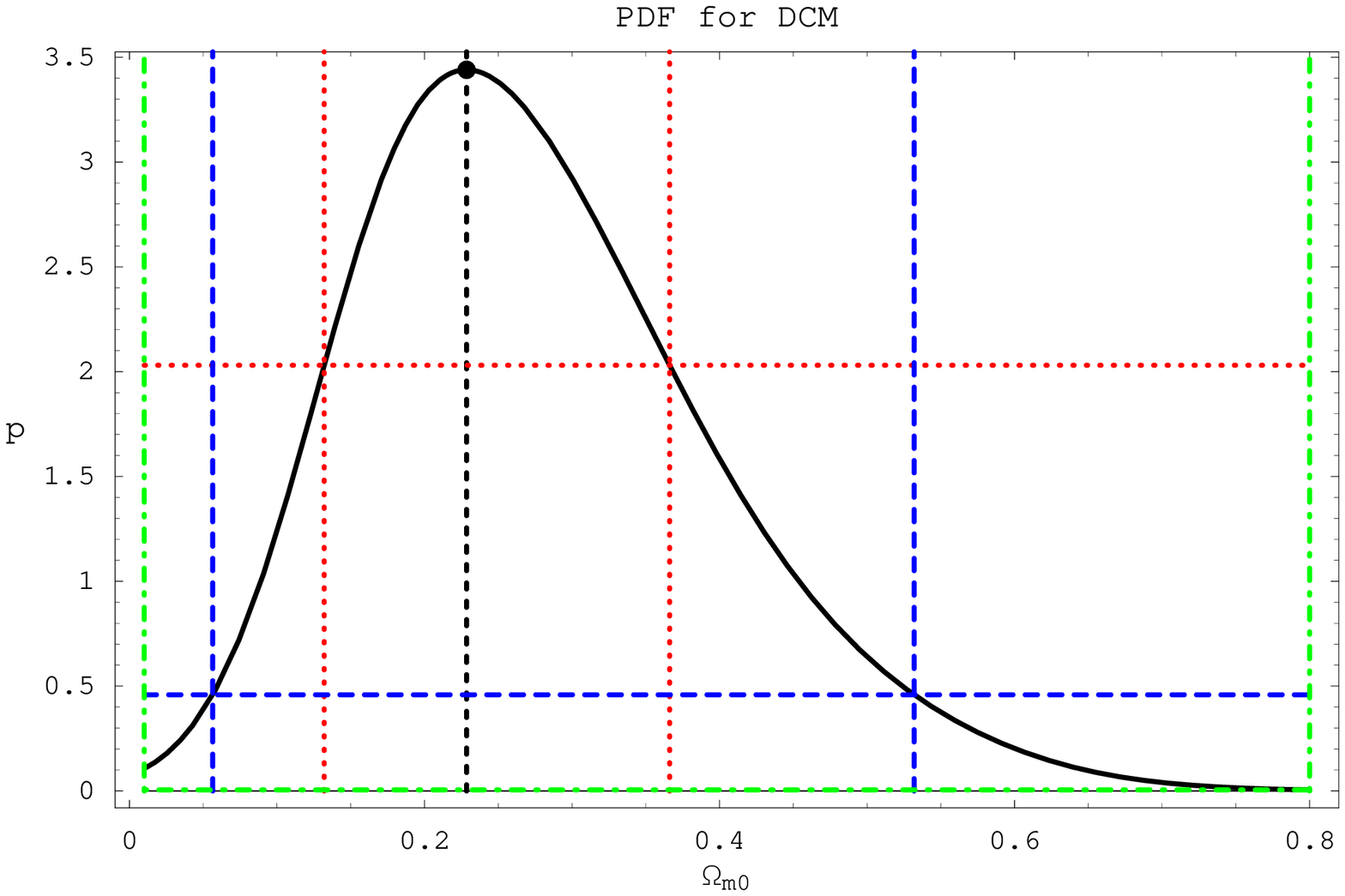}
\end{minipage} \hfill
\caption{{\protect\footnotesize The plots of the PDF as function
of $\protect \Omega_{m0}$ for the two component model, for $k_1
= 0$ and $1.0$. The
joint PDF peak is shown by the large dot, the confidence regions of $1\,%
\protect\sigma $ ($68,27\%$) by the dotted line, the $2\,\protect\sigma $
($95,45\%$) in dashed line and the $3\,\protect\sigma $ ($99,73\%$) in dashed-dotted line.}}
\end{figure}

\begin{figure}[!t]
\begin{minipage}[t]{0.5\linewidth}
\includegraphics[width=\linewidth]{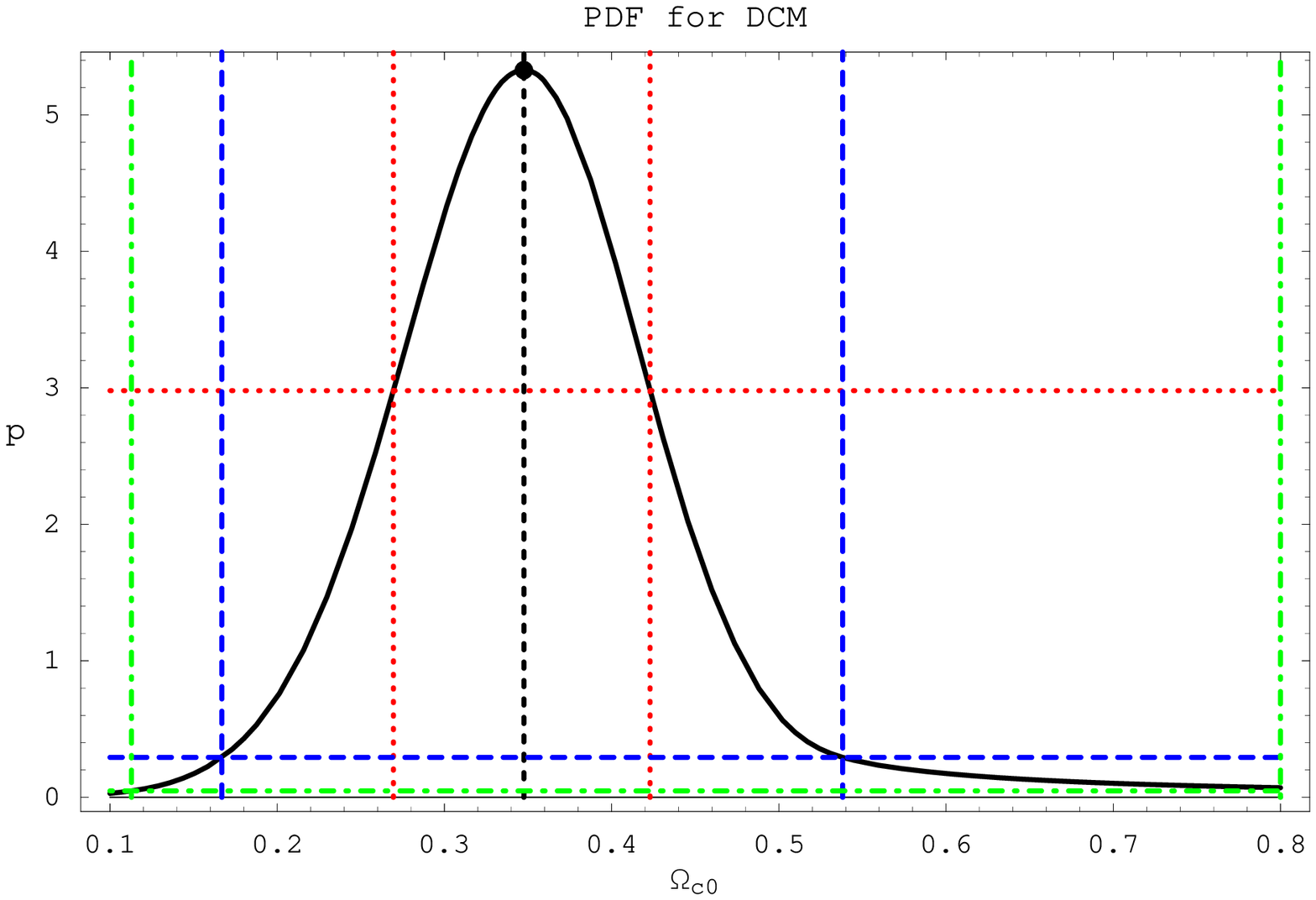}
\end{minipage} \hfill
\begin{minipage}[t]{0.5\linewidth}
\includegraphics[width=\linewidth]{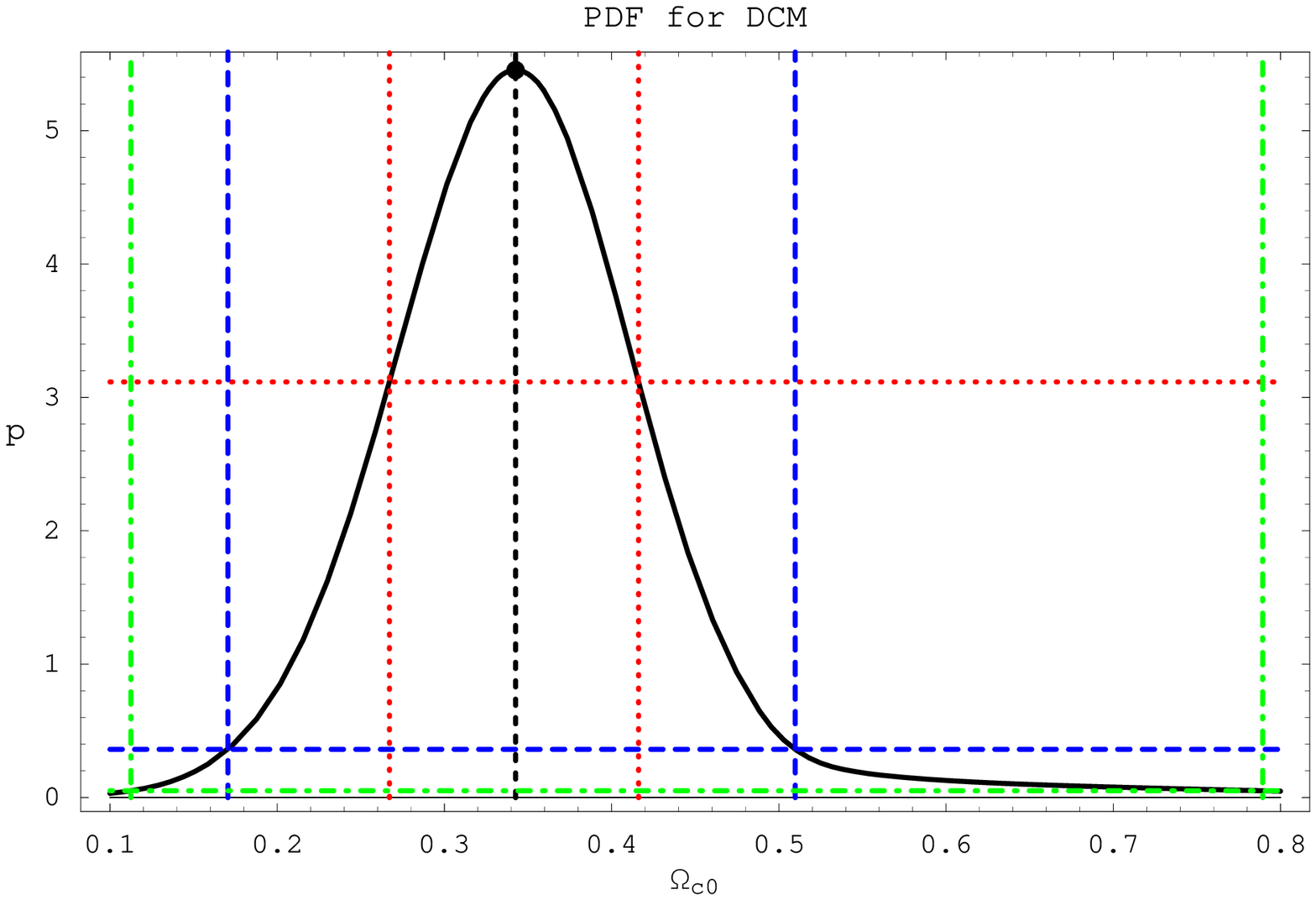}
\end{minipage} \hfill
\caption{{\protect\footnotesize The plots of the PDF as function
of $(\protect \Omega_{c0})$ for the two component model, for $k_1
= 0$ and $1.0$. The
joint PDF peak is shown by the large dot, the confidence regions of $1\,%
\protect\sigma $ ($68,27\%$) by the dotted line, the $2\,\protect\sigma $
($95,45\%$) in dashed line and the $3\,\protect\sigma $ ($99,73\%$) in dashed-dotted line.}}
\end{figure}

\begin{figure}[!t]
\begin{minipage}[t]{0.5\linewidth}
\includegraphics[width=\linewidth]{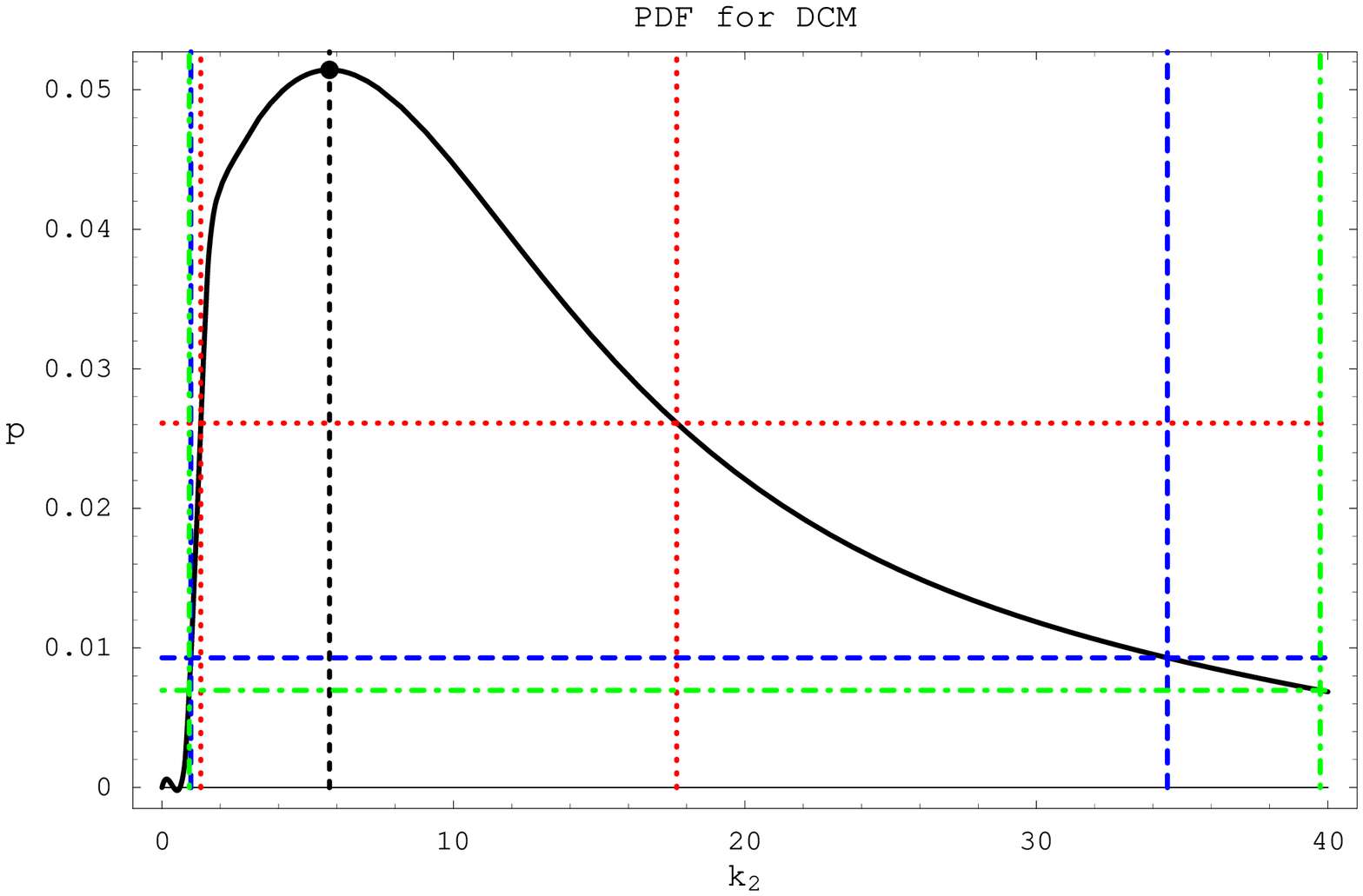}
\end{minipage} \hfill
\begin{minipage}[t]{0.5\linewidth}
\includegraphics[width=\linewidth]{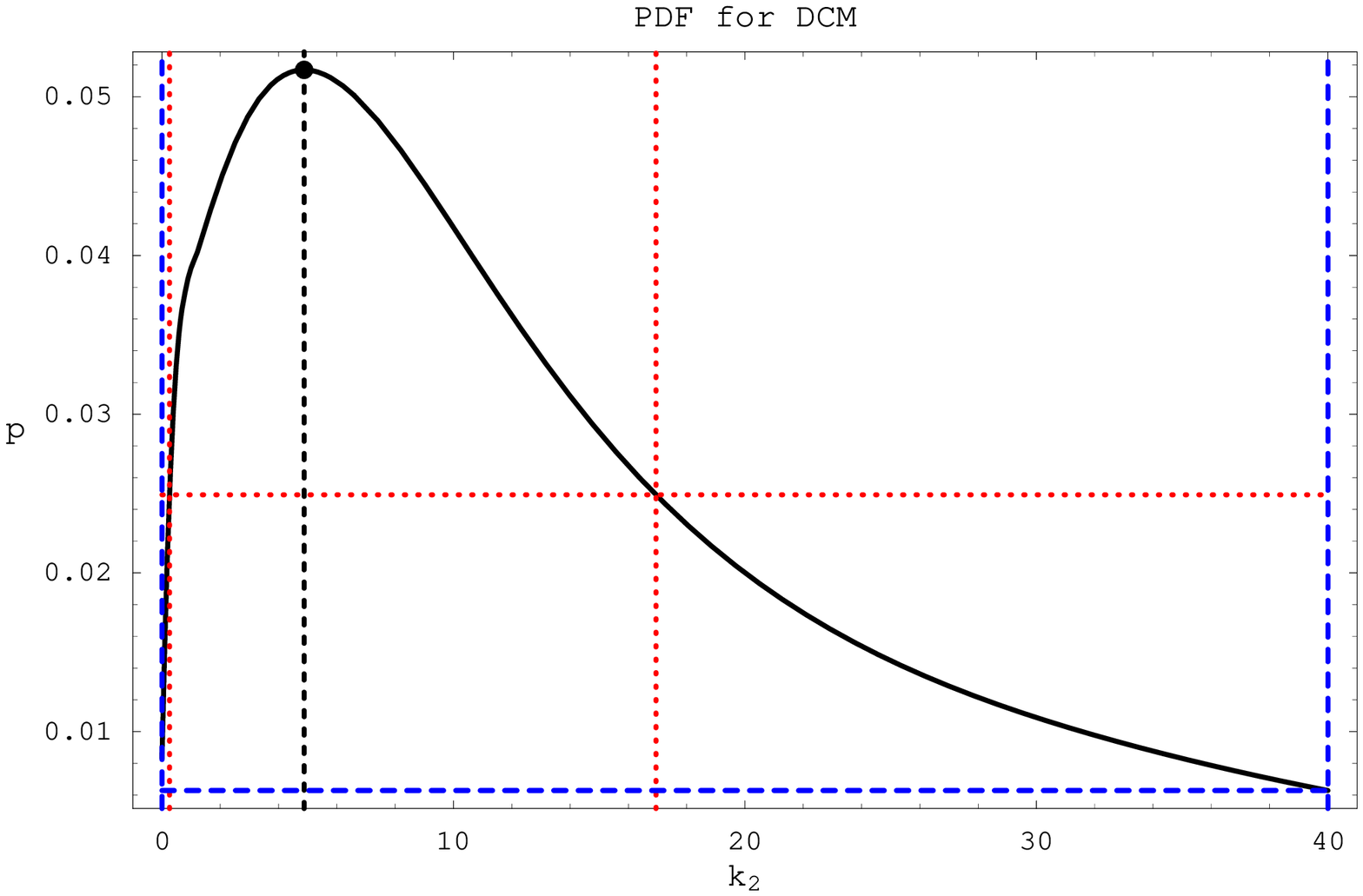}
\end{minipage} \hfill
\caption{{\protect\footnotesize The plots of the PDF as function
of $(\protect k_2)$ for the two component model, for $k_1 = 0$ and $1.0$. The
joint PDF peak is shown by the large dot, the confidence regions of $1\,%
\protect\sigma $ ($68,27\%$) by the dotted line, the $2\,\protect\sigma $
($95,45\%$) in dashed line and the $3\,\protect\sigma $ ($99,73\%$) in dashed-dotted line.}}
\end{figure}

\begin{figure}[!t]
\begin{minipage}[t]{0.5\linewidth}
\includegraphics[width=\linewidth]{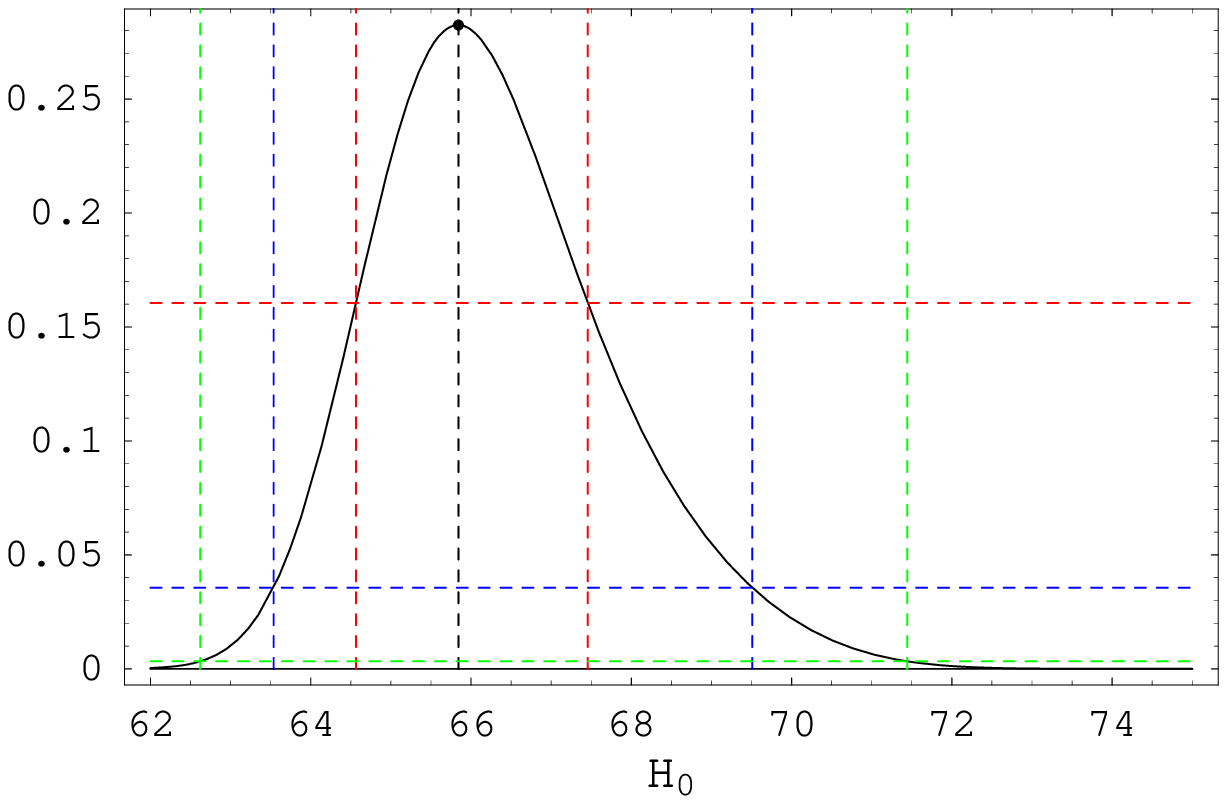}
\end{minipage} \hfill
\begin{minipage}[t]{0.5\linewidth}
\includegraphics[width=\linewidth]{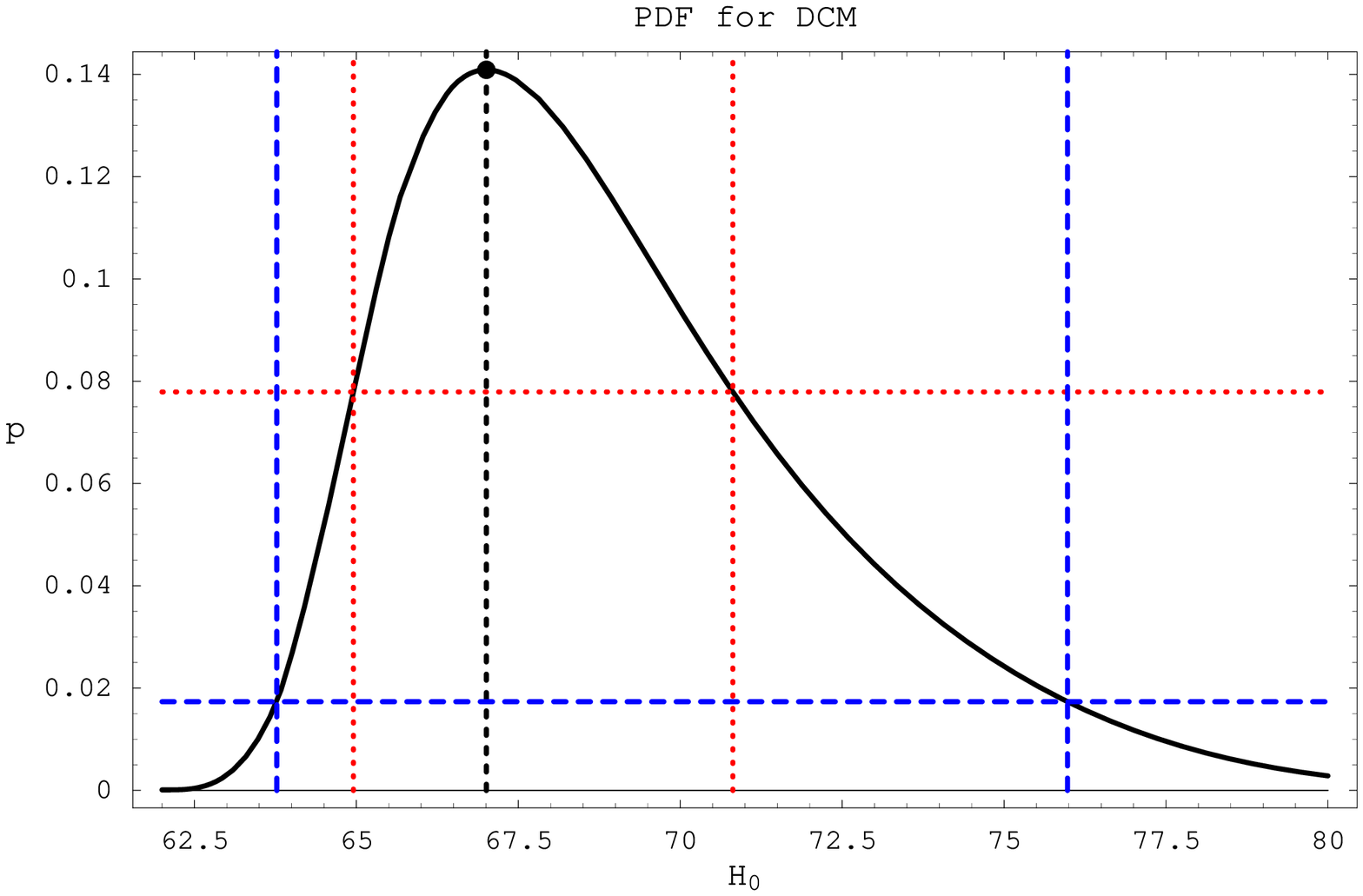}
\end{minipage} \hfill
\caption{{\protect\footnotesize The plots of the PDF as function
of $(\protect h)$ for the two component model, for $k_1 = 0$ and $1.0$. The
joint PDF peak is shown by the large dot, the confidence regions of $1\,%
\protect\sigma $ ($68,27\%$) by the dotted line, the $2\,\protect\sigma $
($95,45\%$) in dashed line and the $3\,\protect\sigma $ ($99,73\%$) in dashed-dotted line.}}
\end{figure}
\par
A more precise estimation of the parameters can be obtained by
evaluating the one-dimensional PDF, by integrating on the three
other parameters. The results are displayed in figures $5$, $6$,
$7$ and $8$, and confirm what has been said based on the
two-dimensional graphics. The error bars at $1\sigma$, $2\sigma$
and $3\sigma$  confidence levels are indicated. For the matter
density parameter, its preferred value varies from $\Omega_{m0} =
0.233$ for $k_1 = 0$ to $\Omega_{m0} = 0.228$, for $k_1 = 1.0$, with
a very limited dispersion. The preferred value for $\Omega_{c0}$
decreases also from $0.347$ to $0.343$. For $H_0$, the preferred
value remains $H_0 = 67.01 km/s\cdot Mpc$ with a small dispersion.
However, for $k_2$, the preferred value varies from $k_2 = 5.75$
for $k_1 = 0$ to $k_2 = 4.88$ for $k_1 = 1.0$. However, the
dispersion is extremely large. Observe that $k_1 = 0$ implies that
our model reduces to a single component dark energy fluid. In this
case, phantom fluids ($k_2 > 1$) are clearly preferred. Hence, an
open universe, dominated by a phantom field, is a generic
prediction of the model. Notice that the age of the universe is
compatible with other astrophysical estimations, remaining around
$13.6\,Gy$.
\par
In order to make a proper comparison, the same analysis for the
$\Lambda$CDM, using the same supernova sample, leads to $\Omega_{m0} =
1.01^{+1.08}_{-0.085}$ and $H_0 = 65.0^{+1.78}_{-1.74}\,km/s\cdot
Mpc$ \cite{colistete}. Hence, the double component fluid model
exploited here predicts a low density universe in strong contrast
with the $\Lambda CDM$ model. The lowest value for the $\chi^2$ is
$1.10$ four the double component model, and $1.11$ for the
$\Lambda CDM$ model. This shows that the double component model is
quite competitive.
\newline
\vspace{1.0cm}
\newline
\begin{table}[t]
\begin{center}
\begin{tabular}{|c|c|c|c|}
\hline
$k_1$&0.0&0.5&1.0 \\[3pt] \hline
$\Omega_{m0}$&$0.233^{+0.303}_{-0.171}$&$0.231^{+0.305}_{-0.172}$&$0.228^{+0.0.304}_{-0.172}$\\[5pt] \hline
$\Omega_{c0}$&$0.347^{+0.191}_{-0.181}$&$0.345^{+0.180}_{-0.178}$&$0.343^{+0.167}_{-0.172}$\\[5pt] \hline
$\Omega_{k0}$&$0.408^{+0.317}_{-0.467}$&$0.412^{+0.317}_{-0.457}$&$0.415^{+0.318}_{-0.439}$ \\[5pt] \hline
$H_0$&$67.01^{+8.80}_{-3.21}$&$67.01^{+8.89}_{-3.24}$&$67.01^{+8.97}_{-3.24}$
\\[3pt] \hline
$k_2$&$5.75^{+28.74}_{-4.75}$&$5.35^{+28.89}_{-4.85}$&$4.88^{+35.12}_{-4.88}$\\[5pt] \hline
$t_0$&$13.64^{+1.20}_{-0.74}$&$13.65^{+1.20}_{-0.75}$&$13.65^{+1.22}_{-.075}$\\[5pt] \hline
\end{tabular}
\end{center}
\caption{Estimated values for $\Omega_{m0}$, $\Omega{c0}$, $\Omega_{k0}$, $H_0$, $k_2$ and $t_0$ for
three different values of $k_1$, at $2\sigma$ level.}
\end{table}

\section{Conclusion}

In this work, we have explored the possibility that the dark
energy has an equation of state given by (\ref{eos}). This
proposal has already been explored in reference \cite{alcaniz1},
but restricting one of the components to behave like a
cosmological constant: a two-component fluid, which is a
"variation" around a cosmological constant, has been analysed.
Here, we alleviate this restriction, but introducing another one:
the linear component can have any barotropic index, but the second
component must vary as the square root of the density. This allows
us to obtain an analytical expression for the evolution of the
Universe, at least for a flat spatial section. This analytic expression reveals that it is possible to
have an asymptotically de Sitter phase, for $p < 0$ and $k_2 < 1$,
even if only $k_2 = 1$ represents the cosmological constant. This
is an intriguing aspect of the model. For $k_2 > 1$ there is
always a big rip.
\par
The restriction in the polytropic factor $\alpha$ seems not to be
so relevant in view of the results of reference \cite{alcaniz1}:
if the second component obeys a polytropic power law, there is a
strong degeneracy on the polytropic factor, and almost any value
of the power is allowed.
\par
Our results
indicate an open universe with a matter density parameter around $\Omega_{m0}
\sim 0.3$, with a similar estimation for the dark energy component.
The $\Lambda CDM$ model favours a closed universe. On the other hand, the predicted
value for the Hubble parameter is more consistent with other observational tests, like CMB, that
is, $H_0 \sim 67\,km/s\cdot Mpc$
\cite{spergel}. For the barotropic index in the two-component
fluid $k_2$, the results indicate that $k_2
> 1$ is highly favoured. The dispersion is very high, but tends still to favour a phantom scenario. As
the component index $k_1$ increases, the preferred value of $k_2$
decreases slightly.
\par
It must be remarked that the model described here exhibits a
$\chi^2$ slightly smaller than for the $\Lambda CDM$ model. This
shows that the model is quite competitive. In our opinion, even if
a phantom scenario is clearly preferred, the fact of predicting a
low density universe is interesting in its own, mainly when
compared with the analysis of clustering of matter in the
universe.

{\bf Acknowledgments:} We thank the anonymous referee who has suggested us, besides other remarks,
to consider the non-flat case, what has led to new interesting results with respect to the previous version of
the paper.
We thank also CNPq (Brazil) and CAPES (Brazil)
for partial financial support. F.C. and J.F.V.R. thank also FAPERJ (Brazil) for partial financial
support. We thank Roberto Colistete Jr. and Martin Makler for their
critical remarks.


\begin{thebibliography}{90}

\bibitem{SN}
A.G. Riess et al., Astron. J. {\bf 116}, 1009(1998); S. Perlmutter et
al., Astrophys. J. {\bf 517}, 565(1999).

\bibitem{tonry} J.L. Tonry et al, Astrophys. J. {\bf 594}, 1(2003).

\bibitem{riess} A.G. Riess, Astrophys. J. {\bf 607}, 665(2004).

\bibitem{astier} P. Astier et al., Astron. Astrophys. {\bf 447}, 31(2006).

\bibitem{verde} L. Verde et al., Astrophys. J. Suppl. {\bf 148},
195(2003).

\bibitem{tegmark1} M. Tegmark et al., Astrophys. J. {\bf 606},
702(2004).

\bibitem{colistete} R. Colistete Jr, J. C. Fabris, S.V.B. Gon\c{c}alves and P.E. de Souza, Int. J. Mod. Phys. {\bf D13}, 669(2004);
R. Colistete Jr., J. C. Fabris and S.V.B. Gon\c{c}alves, Int. J.
Mod. Phys. {\bf D14}, 775(2005); R. Colistete Jr. and J. C.
Fabris, Class. Quant. Grav. {\bf 22}, 2813(2005).

\bibitem{tegmark2} M. Tegmark et al., Phys. Rev. {\bf D69}, 103501(2004).

\bibitem{sahni} V. Sahni, Lect. Notes Phys. {\bf 653}, 141(2004).

\bibitem{fantasma} S. Hannestad and E. Mortsell, JCAP {\bf 0409}, 001 (2004); U. Alam, V. Sahni, T.D. Saini and A.A. Starobinsky,
Mon. Not. R. Astron. Soc. {\bf 354}, 275 (2004); S.W. Allen et al., Mon. Not. R. Astron. Soc. {\bf 353}, 457 (2004).

\bibitem{pad} H.K. Jassal, J.S. Bagla and T. Padmanabhan, {\it The vanishing of phantom menace}, astro-ph/0601389.

\bibitem{chaplygin} A.Yu. Kamenschik, U. Moschella and V. Pasquier, Phys. Lett. {\bf B511}, 265 (2001); J.C. Fabris, S.V.B. Gon\c{c}alves and
P.E. de Souza, Gen. Rel. Grav. {\bf 34}, 53 (2002); N. Bilic, G.B. Tupper and R.D. Viollier, Phys. Lett. {\bf 535}, 17 (2002);
M.C. Bento, O. Bertolami and A.A. Sen, Phys. Rev. {\bf D66}, 043507 (2002).

\bibitem{indianos} S. Mukherjee, B.C. Paul, N.K. Dadhich, S.D. Maharaj and A. Beesham, {\it Emergent universes with exotic matter},
gr-qc/0605134.

\bibitem{benaoum} H.B. Benaoum, {\it Accelerated universe from modified Chaplygin gas and tachyonic fluid}, hep-th/0205140.

\bibitem{chakra1} U. Debnath, A. Banerjee and S. Chakraborty, Class. Quant. Grav. {\bf 21}, 5609 (2004).

\bibitem{chakra2} W. Chakraborty and U. Debnath, {\it Is the modified Chaplygin gas along with barotropic fluid responsible for
acceleration of the universe?}, gr-qc/0611094.

\bibitem{alcaniz1} J.S. Alcaniz and H. Stefancic, {\it "Expansion" around the vacuum: how far can we go
from $\Lambda$?}, astro-ph/0512622.

\bibitem{jerome} J.C. Fabris and J. Martin, Phys. Rev. {\bf D55}, 5205 (1997).

\bibitem{weinberg} S. Weinberg, {\bf Gravitation and cosmology}, Wiley, New York( 1972).

\bibitem{coles} P. Coles and F. Lucchin, {\bf Cosmology}, Wiley,
New York (1995).

\bibitem{betocs}  R. Colistete Jr., \textit{\textbf{B}ay\textbf{E}sian
\textbf{T}ools for \textbf{O}bservational \textbf{C}osmology using
\textbf{S}Ne Ia (\textbf{BETOCS})}, available on the Internet site
http://www.RobertoColistete.net/BETOCS, (2006).

\bibitem{wmap} D.N. Spergel et al., {\it Wilkinson microwave anisotropy probe (WMAP) three years results:
implications to cosmology}, astro-ph/0603449.

\bibitem{spergel} D.N. Spergel et al., Astrophys. J. Suppl. {\bf 148}, 175(2003).

\end{thebibliography}
\end{document}